\documentclass[twocolumn,nofootinbib,amsmath,amssymb,aps,prd,superscriptaddress]{revtex4-1}

\usepackage{graphicx}
\usepackage[caption=false]{subfig}

\usepackage{amsmath,amssymb}
\usepackage{amsfonts,amssymb,mathrsfs}

\usepackage[bookmarks=false,pdfstartview=FitH]{hyperref}
\usepackage[all]{hypcap}
\usepackage{natbib}

\def\be{\begin{equation}}
\def\ee{\end{equation}}
\def\nn{\nonumber}
\def\f{\frac}
\def\tf{\tfrac}
\def\pl{{\rm Pl}}
\def\lp{\ell_\pl}

\def\de{\delta}
\def\ep{\epsilon}

\def\la{\lambda}
\def\om{\omega}

\def\mR{\mathcal{R}}

\usepackage{color}

\begin{document}

\pagestyle{plain}

\title{Modified dispersion relations, inflation and scale invariance}

\author{Stefano Bianco} \email{stefano.bianco@aei.mpg.de}
\affiliation{Max Planck Institute for Gravitational Physics (Albert Einstein Institute),\\
Am M\"uhlenberg 1, 14476 Golm, Germany, EU}

\author{Victor Nicolai Friedhoff} \email{nicolai.friedhoff@aei.mpg.de}
\affiliation{Max Planck Institute for Gravitational Physics (Albert Einstein Institute),\\
Am M\"uhlenberg 1, 14476 Golm, Germany, EU}

\author{Edward Wilson-Ewing} \email{edward.wilson-ewing@unb.ca}
\affiliation{Max Planck Institute for Gravitational Physics (Albert Einstein Institute),\\
Am M\"uhlenberg 1, 14476 Golm, Germany, EU}
\affiliation{Department of Mathematics and Statistics, University of New Brunswick, \\
Fredericton, NB, Canada E3B 5A3
}

\begin{abstract}

For a certain type of modified dispersion relations, the vacuum quantum state for very short wavelength cosmological perturbations is scale-invariant and it has been suggested that this may be the source of the scale-invariance observed in the temperature anisotropies in the cosmic microwave background.  We point out that for this scenario to be possible, it is necessary to redshift these short wavelength modes to cosmological scales in such a way that the scale-invariance is not lost.  This requires non-trivial background dynamics before the onset of standard radiation-dominated cosmology; we demonstrate that one possible solution is inflation with a sufficiently large Hubble rate, for this slow roll is not necessary.  In addition, we also show that if the slow-roll condition is added to inflation with a large Hubble rate, then for any power law modified dispersion relation quantum vacuum fluctuations become nearly scale-invariant when they exit the Hubble radius.

\end{abstract}

\maketitle

\section{Introduction}
\label{s.intro}

According to many inflationary models, at the beginning of inflation the entire observable universe today was initially the size of the Planck volume, or perhaps even smaller.  Since quantum gravity effects are typically expected to become important near the Planck scale, this suggests that quantum gravity effects may have been important at the onset of inflation.  The potential importance of as yet unknown quantum gravity effects in inflation is known as the trans-Planckian problem of inflation.

One potential quantum gravity effect that could be important for trans-Planckian cosmological perturbations is the possible dimensional reduction of space-time to two dimensions in the ultraviolet \cite{Carlip:2016qrb}.  In a large number of candidate theories of quantum gravity, some measure of dimensionality (whether the geometric,  spectral, thermodynamic or Myrheim-Meyer dimension) runs to 2 near the Planck scale, including: high temperature string theory \cite{Atick:1988si}, causal dynamical triangulations \cite{Ambjorn:2005db}, asymptotically safe gravity \cite{Litim:2003vp, Lauscher:2005qz}, Ho{\v{r}}ava-Lifshitz gravity \cite{Horava:2009if}, space-time non-commutativity \cite{Benedetti:2008gu, Alesci:2011cg}, loop quantum gravity and spin foam models \cite{Modesto:2008jz, Magliaro:2009if, Calcagni:2014cza}, multi-fractional space-times \cite{Calcagni:2010pa}, and causal sets \cite{Carlip:2015mra, Belenchia:2015aia}.

Interestingly, it appears possible to capture many aspects of the running of the spectral dimension to 2 by introducing modifications to the dispersion relation of the type \cite{Horava:2009if, Sotiriou:2011aa}
\be \label{mod-disp}
E^2 = k^2 + \sigma^4 k^6,
\ee
for massless particles, where $\sigma$ is a length scale which is typically assumed to be of the order of or slightly larger than the Planck length $\lp$.  (The spectral dimension is determined by calculating the scaling of the average return probability of a diffusion process, see \cite{Ambjorn:2005db} for details.  See \cite{Amelino-Camelia:2013gna} for complementary results relating \eqref{mod-disp} to an additional form of dimensional reduction to 2 in the ultraviolet.)  Importantly, for a dimensional reduction to 2 spectral dimensions in the ultraviolet, the correction terms truncate at the $k^6$ term \cite{Sotiriou:2011aa}.  In principle, there could be other modifications in \eqref{mod-disp} of lower order in $k$ than $k^6$---although any potential prefactor to the $k^3$ term is strongly constrained by observations \cite{Ackermann:2009aa}---but in any case the term that is of most interest in the cosmological context is the $\sigma^4 k^6$ term.  In addition, some estimates based on backreaction effects in loop quantum cosmology also suggest that the $k^6$ term may be the most relevant one in modified dispersion relations \cite{Bodendorfer:2016uat}.

Note that a modified dispersion relation like \eqref{mod-disp} may result from a theory which explicitly breaks Lorentz invariance near the Planck scale, like Ho{\v{r}}ava-Lifshitz gravity for example, but can also arise due to deformations of the Lorentz symmetries \cite{AmelinoCamelia:2000mn, KowalskiGlikman:2001gp, Magueijo:2002am}.  While many Lorentz-violating theories are ruled out by low energy quantum field theory---for example, all theories that predict a Planck-scale discreteness in a preferred frame \cite{Collins:2004bp}---theories that merely deform the Lorentz symmetries are not immediately ruled out by these constraints.

Modified dispersion relations have previously been considered for a number of phenomenological studies in quantum gravity \cite{AmelinoCamelia:1997gz, Gambini:1998it, Alfaro:1999wd} and cosmology \cite{Martin:2000xs, Brandenberger:2000wr, Niemeyer:2000eh, KowalskiGlikman:2000dz, Lemoine:2001ar}.  Perhaps most strikingly, for modified dispersion relations that truncate at the $k^6$ term like \eqref{mod-disp}, the vacuum quantum state for the shortest wavelengths of cosmological perturbations is known to be scale-invariant \cite{Magueijo:2008yk}.  This immediately suggests the possibility that modified dispersion relations could be the source of the observed scale-invariance in the cosmic microwave background (CMB) rather than slow-roll inflation \cite{Magueijo:2008yk, Amelino-Camelia:2013tla}.

Despite the appeal of this idea, there are two challenges that must be addressed for this scenario to be possible.

First, since the cosmological perturbations are only scale-invariant for the wavelengths where the $k^6$ term dominates the modified dispersion relation (i.e., for modes with wavelengths comparable to or smaller than $\sigma$, where $\sigma$ should be of the order $\sim 10^4 \lp$ to obtain the correct amplitude for the power spectrum, see below), it is necessary that these modes be redshifted to cosmological scales by the expansion of the universe.  Since the longest-wavelength perturbations observed in the CMB are $\sim 10^{60} \lp$, a total redshift factor of $\sim 10^{56}$ is needed.  With the standard cosmological model without inflation providing a redshift factor of $\sim 10^{30}$, it follows that this scenario requires an additional earlier era of expansion during which the scale factor increases by a factor of (at a minimum) $\sim 10^{26}$.

Second, as already pointed out in \cite{Cai:2009hc}, these perturbations must exit their sound horizon and freeze at a time when the $k^6$ term still dominates in the modified dispersion relation \eqref{mod-disp}, otherwise it is known that the scale-invariance is lost in short-wavelength perturbations when the $k^2$ term in the dispersion relation becomes important \cite{Martin:2000xs, Brandenberger:2000wr, Niemeyer:2001qe} (although of course cosmological scenarios like inflation can restore scale-invariance at a later time via a different mechanism than modified dispersion relations); note that another reason to require that the perturbations freeze is that this is needed to generate the phase coherence observed in the acoustic peaks of the CMB scalar power spectrum \cite{Dodelson:2003ip}.

These two challenges can be met by assuming that there exist some non-trivial dynamics before the onset of the standard cosmological dynamics.  One such example that we will consider here is inflation with a large Hubble rate.  Of course, it may be possible to find other early-time dynamics for the background FLRW space-time that also successfully address these two challenges.

We begin in Sec.~\ref{s.mod} with a brief self-contained review of the calculation given in \cite{Magueijo:2008yk} demonstrating that modified dispersion relations of the form \eqref{mod-disp} give, for short wavelength cosmological perturbations, a vacuum quantum state that is already scale-invariant.  In Sec.~\ref{s.infl} inflation with a large Hubble rate $H > \sigma^{-1}$ is given as an example of a simple cosmological scenario that redshifts these perturbations to cosmological scales, as required, without losing scale-invariance.

Then, in Sec.~\ref{s.obs} we show that (assuming quantum vacuum initial conditions) slow-roll inflation with a sufficiently large Hubble rate in fact generates a nearly scale-invariant power spectrum of cosmological perturbations no matter the modified dispersion relation.  This is the interesting converse of the result obtained in \cite{Magueijo:2008yk} that the modified dispersion relation \eqref{mod-disp} gives scale-invariant perturbations at short scales no matter the cosmological background; here it is slow-roll inflation that generates scale-invariant perturbations no matter the modified dispersion relation.  We end with a discussion in Sec.~\ref{s.disc}.

Finally, note that the equations of motion proposed for cosmological perturbations with modified dispersion relations vary in the literature.  In some cases, it is the equation of motion for cosmological perturbations, including the backreaction of perturbations in the matter fields on the space-time at linear order, that are transformed to include the modified dispersion relations \cite{Martin:2000xs, Brandenberger:2000wr, Niemeyer:2001qe}, while in other cases it is the equation of motion for a test field on a cosmological background that is modified \cite{Magueijo:2008yk, Amelino-Camelia:2013tla, Cai:2009hc}.  In this paper, we will follow \cite{Martin:2000xs, Brandenberger:2000wr, Niemeyer:2001qe}.  This choice makes no difference in Sec.~\ref{s.infl} where a constant equation of state for the matter field is assumed, in which case the equations of motion proposed in \cite{Martin:2000xs, Brandenberger:2000wr, Niemeyer:2001qe} and \cite{Magueijo:2008yk, Amelino-Camelia:2013tla, Cai:2009hc} become identical.  (The relation between the perturbation variables used in these papers and the co-moving curvature perturbation is also different, but for the case that the equation of state of the matter field is constant, these relations become identical also, up to an overall numerical prefactor.)  On the other hand, the predictions for slow-roll inflation presented in Sec.~\ref{s.obs} will depend on the choice of the equation of motion, as in this case the equation of state is not constant.


\section{Scale-Invariance from Modified Dispersion Relations}
\label{s.mod}

In standard cosmological perturbation theory for general relativity coupled to a scalar field $\phi$, the dynamics of scalar perturbations are governed by the Mukhanov-Sasaki equation,
\be
v_k'' + k^2 v_k - \f{z''}{z} v_k = 0,
\ee
where $v_k$ are Fourier modes of the Mukhanov-Sasaki variable $v$ related to the gauge-invariant co-moving curvature perturbation $\mR$ by $\mR = v / z$, and $z = a \dot{\phi} / H$ is a function of the background space-time with $a$ being the scale factor and $H = \dot{a}/a$ is the Hubble rate.  (If the matter field is a perfect fluid rather than a scalar field, then $z = a \sqrt{\rho + P \, } / c_s H$, where $\rho, P,$ and $c_s$ are the energy density, pressure and sound speed of the fluid, respectively.)  Here primes denote derivatives with respect to the conformal time $\tau$ while dots denote derivatives with respect to the proper time $t$; they are related by $f' = a \dot f$.  For an introduction to cosmological perturbation theory, see, e.g., \cite{Mukhanov:1990me, Mukhanov:2005sc}.

Allowing for modified dispersion relations of the form \eqref{mod-disp} gives the modified Mukhanov-Sasaki equation \cite{Martin:2000xs}
\be \label{ms-mod}
v_k'' + \left(k^2 + \f{\sigma^4 k^6}{a^4} \right) v_k
- \f{z''}{z} v_k = 0.
\ee
(For an explicit example of how modified dispersion relations appear in the equations of motion for cosmological perturbations in Ho{\v{r}}ava-Lifshitz gravity, see \cite{Takahashi:2009wc, Calcagni:2009ar}.)

Now, for the case considered here where the relevant modification to the dispersion relation is the $k^6$ term---which is argued to capture some important aspects of the possible dimensional reduction of space-time to 2 dimensions in the ultraviolet \cite{Horava:2009if, Sotiriou:2011aa}---the vacuum quantum state for $v_k$ is scale-invariant \cite{Magueijo:2008yk}.  This follows from the result that for the shortest wavelength modes, the above equation is well approximated by
\be \label{diff-k6}
v_k'' + \f{\sigma^4 k^6}{a^4} v_k = 0,
\ee
which is a simple harmonic oscillator with the time-dependent frequency $\om_k = \sigma^2 k^3 / a^2$.  If the condition $|\om_k' / \om_k^2| \ll 1$ holds (i.e., that $a'/a \ll \om_k$), then the WKB solution
\be \label{vac}
v_k = \f{A_k}{\sqrt{\om_k}} \, e^{-i \int d\tau \, \om_k}
+ \f{B_k}{\sqrt{\om_k}} \, e^{i \int d\tau \, \om_k}
\ee
provides a good approximation to the solution of \eqref{diff-k6}.  Imposing that $v_k$ initially be in the vacuum quantum state implies that $A_k = \sqrt{\hbar / 2}$ and $B_k = 0$ \cite{Martin:2000xs}.

An immediate consequence of this result is that the power spectrum of the co-moving curvature perturbation,
\be
\Delta_\mR^2(k) = \f{k^3}{2 \pi^2} |\mR_k|^2 = \f{\hbar}{4 \pi^2 \sigma^2} \cdot \f{a^2}{z^2},
\ee
is independent of $k$ and hence is scale-invariant.

Furthermore, the $(a/z)^2$ term can be simplified if the dynamics of the background is known.  Specifically, given the equation of state $w = P / \rho$ relating the pressure $P = \dot{\phi}^2 / 2 - V(\phi)$ to the energy density $\rho = \dot{\phi}^2 / 2 + V(\phi)$ of the scalar field, $z$ can be rewritten (using the Friedmann equation) as $z = a \sqrt{3 (1+w) / 8 \pi G}$, and then
\be
\Delta_\mR^2(k) = \f{2 G \hbar}{3 \pi (1+w) \sigma^2}.
\ee
Since $w$ is typically of order 1 (except for the case of slow-roll inflation where $w$ is close to $-1$, this possibility is considered in Sec.~\ref{s.obs}), for the choice $\sigma \sim 10^4 \sqrt{G\hbar}$, then not only is the power spectrum of the scalar perturbations scale-invariant, but its amplitude also matches what has been observed in the cosmic microwave background \cite{Ade:2015xua}.

Of course, a slight red tilt has also been observed in the power spectrum of the CMB.  In this scenario, it is possible to obtain a slight red tilt by considering a dispersion relation where the modification to the dispersion relation is a term of the type $k^{6 - \de}$ with $0 < \de \ll 1$, or by considering the situation where there is a long transient between the $k^6$ term and the $k^2$ where the modified dispersion relation can be approximated by $k^{6 - \de}$ \cite{Amelino-Camelia:2013tla}.

A similar calculation can be performed for tensor modes with similar results \cite{Amelino-Camelia:2013tla}, although in order to obtain a sufficiently small tensor-to-scalar ratio it is necessary to assume that the length scale $\sigma_T$ in the modification to the dispersion relation for tensor modes $\sigma_T^4 k^6 / a^4$ is much smaller than $\sigma$.  Whether this difference in the modifications to the dispersion relations of tensor and scalar modes can be explained naturally by quantum gravity remains an open question.  We will discuss another possibility to obtain a red tilt for the spectrum of scalar perturbations and a small tensor-to-scalar ratio in Sec.~\ref{s.obs}.

Now the question is whether this effect from modified dispersion relations could in fact be the ultimate source of the temperature anisotropies observed in the cosmic microwave background.  For this to be possible, there are two requirements: (i) first, these modes that are scale-invariant have a very short wavelength of the order of $\sigma$ or less and therefore must be redshifted to cosmological scales \cite{Piao:2009ax}, which requires a new cosmological era before radiation-domination during which the scale factor increases by a factor of $\sim 10^{26}$, and (ii) these scale-invariant modes must exit the horizon before $k^2$ becomes comparable to $\sigma^4 k^6 / a^4$, otherwise the dominant term in \eqref{ms-mod} will no longer be the $k^6$ term and scale-invariance will be lost, as has been studied in some detail in \cite{Martin:2000xs, Brandenberger:2000wr, Niemeyer:2001qe} and will be reviewed in Sec.~\ref{s.infl} below.

One simple cosmological scenario that can easily address both of these requirements is inflation with a very large Hubble rate.  While there may be other possibilities, for both of the above requirements to be met any cosmological scenario must generate a Hubble rate that remains above or at the inverse length scale $\sigma^{-1}$ for at least 8 e-folds (the number of e-folds of Fourier wavenumbers observed in the CMB today), and must provide at least $\sim 70$ additional e-folds of expansion.  Inflation with a sufficiently large Hubble rate is probably the simplest cosmological scenario that meets these requirements.  In fact, for the case that the dynamics of the universe is dominated by a matter field with a constant equation of state, inflation is required for there to be a sufficient number of e-folds (under the reasonable assumption that the length-scale $\sigma$ is not more than $\sim 24$ orders of magnitude larger than the Planck scale) \cite{Piao:2009ax}.

\section{Near-Planck-Scale Inflation}
\label{s.infl}

Inflation occurs when the equation of state of the matter field is less than $w < -1/3$.  Here we will assume that the equation of state is constant in time in order to simplify the calculations, but the results in this section can be generalized in a straightforward fashion in order to allow for a dynamical $w$.

For a constant equation of state $w < -1/3$, the scale factor behaves as
\be \label{scale}
a(\tau) = a_o \, (-\tau)^{\f{2}{1+3w}},
\ee
where $-\infty < \tau < 0$ for an expanding universe.

Given this form of the scale factor (and using the relation $z''/z = a''/a$ that is valid when the equation of state is constant), the Mukhanov-Sasaki equation with a modified dispersion relation (in the case where this modified dispersion relation has only an additional $k^6$ term) given in \eqref{ms-mod} has the form
\be \label{ms-w}
v_k'' + \left( k^2 + \f{\sigma^4 k^6}{a_o^4 |\tau|^{8/(1+3w)}} \right) v_k
- \f{2(1-3w)}{(1+3w)^2 \tau^2} v_k = 0.
\ee
Since $(1+3w) < 0$ during inflation, at the very early times $\tau \to -\infty$ the $k^6$ term dominates and at this time the vacuum quantum of the Fourier modes will be given by \eqref{vac} (with $\om_k = \sigma^2 k^3 / a^2$) which is scale-invariant.

At late times, when $\tau \to 0$, it is the $z''/z$ which goes as $\tau^{-2}$ which will dominate.  Therefore, for inflation, the modes originate deep inside the Hubble radius where the modifications to the dispersion relation are important, and later exit the horizon and freeze.  (It was incorrectly claimed in \cite{Amelino-Camelia:2013tla} due to a sign error that in an expanding inflationary background these modes do not exit the horizon.  As is clear from \eqref{ms-w}, the modes will indeed exit the horizon in an expanding inflationary background; see also \cite{Cai:2009hc}.)

Note that for scale-invariance to be preserved, the lower order terms in $k$ must not become the dominant term in \eqref{ms-w} before horizon exit.  To see this, consider the case where the $k^2$ term does become relevant.  Using the WKB approximation again, for Fourier modes that are well inside the Hubble radius in which case the $z''/z$ term is negligible, assuming vacuum initial conditions the solution is
\be
v_k = \sqrt \f{\hbar}{2 \tilde{\om}_k} \, e^{-i \int d\tau \, \tilde{\om}_k},
\ee
where $\tilde{\om}_k^2 = k^2 + \sigma^4 k^6 / a^4$.  From this, it is clear that if the $k^2$ term is larger than (or comparable to) the $k^6$ term, the $v_k$ perturbation is no longer scale-invariant.

On the other hand, if the Fourier modes exit the Hubble radius while the $k^6$ term is still dominant, then the scale-invariance is preserved.  This can be seen by solving the Mukhanov-Sasaki equation for super-horizon Fourier modes, which in this limit simplifies to
\be
v_k'' - \f{a''}{a} v_k = 0,
\ee
(again using the relation $z''/z = a''/a$ when the equation of state $w$ is constant), with the solution
\be
v_k = C_k a + D_k a \int \f{d\tau}{a^2}.
\ee
In an expanding inflationary universe, the dominant term will be the $C_k a$ term, as the second term will decay very rapidly compared to the first term as the space-time expands.  The scale-dependence of $v_k$ is contained in the prefactor $C_k$, which can be determined by imposing continuity in $v_k$ and $v_k'$ at the time of horizon-crossing.  Since the time-dependence of $v_k$ inside the horizon (neglecting the unimportant phase) is also given by $a$, it follows that $C_k = \sqrt{ \hbar / 2 \sigma^2 k^3 }$ (recall that here we are considering the case where the $k^6$ term dominates before horizon-crossing).  The result for the super-horizon modes,
\be
v_k = \f{\sqrt \hbar}{\sqrt 2 \sigma k^{3/2}} a,
\ee
(disregarding the second term which rapidly decays) is clearly scale-invariant.

So, for the scale-invariance generated by the modified dispersion relations to be preserved as the space-time expands, it is necessary that the $z''/z$ term become the dominant term in the modified Mukhanov-Sasaki equation well before the $k^2$ term becomes comparable to the $k^6$ term.

This is a constraint on the Hubble rate of the background space-time: it is only for inflationary models where the Hubble rate is sufficiently large that the above condition will be satisfied.  At horizon-crossing, the above condition is
\be \label{ineq}
\f{\sigma^4 k^6}{a^4} = \f{z''}{z} \gg k^2.
\ee
The inequality $\sigma^4 k^4 / a^4 \gg 1$ implies that the physical wavelength $\la_{phys} = a / k$ of the relevant Fourier modes of the cosmological perturbations must be much smaller than the length scale $\sigma$,
\be
\la_{phys} \ll \sigma.
\ee
Furthermore, for the scale factor \eqref{scale} the Hubble rate is
\be
H = \f{a'}{a^2} = \f{2}{|1+3w| \cdot a(\tau) \cdot |\tau|},
\ee
and the $z''/z$ term can be rewritten as
\be
\f{z''}{z} = \f{(1 - 3w)a(\tau)^2}{2} \, H^2.
\ee
From this, the inequality $z''/z \gg k^2$ can be rewritten as
\be
H^2 \gg \f{2 k^2}{(1 - 3w) a^2} = \f{2}{(1 - 3w) \la_{phys}^2},
\ee
and since $\la_{phys} \ll \sigma$, this implies that
\be \label{bound-H}
H \gg \sqrt \f{2}{1 - 3w} \, \sigma^{-1},
\ee
where the Hubble rate is to be evaluated at the horizon-crossing time when $z''/z = \sigma^4 k^6 / a^4$ (which of course depends on the Fourier wavenumber $k$).

If $\sigma \sim 10^{4} \sqrt{G \hbar}$ in order to generate the observed amplitude of the scalar perturbations, this condition requires that the Hubble rate at the crossing time be much greater than $H \gg 10^{-4} / \sqrt{G \hbar}$ (up to some prefactor of order 1 that depends on $w$).

Note also that if there are additional terms in the modified dispersion relation, they must also always be smaller than the first two terms in \eqref{ineq} at horizon crossing.  These will provide additional constraints that, depending on the amplitude of terms, may lead to an even greater lower bound for the Hubble rate.

An important point is that the above calculations do not constrain the number of e-folds of the inflationary era, and therefore it is possible for the scale factor of the universe to increase by a factor $\sim 10^{26}$ (or more), as required for the initially Planck-length (or more precisely, with a wavelength $\sim \sigma$) cosmological perturbations to reach cosmological scales today.

The main conclusion of this section is that inflation can redshift the cosmological perturbations by the required factor $\gtrsim 10^{26}$, while preserving the scale-invariance generated by modified dispersion relations---and note that slow-roll conditions are not necessary, any $w<-1/3$ will do---but that for this to be possible the Hubble rate at the horizon-crossing time of all of the modes of observational interest must have been very large, $H \gg 10^{-4} / \sqrt{G \hbar}$.  Related to this point, note that the amplitude of the co-moving curvature perturbation $\mR_k = \sqrt{4 \pi G \hbar / 3 \sigma^2 (1+w)} \cdot k^{-3/2}$ is independent of the Hubble rate (unlike in the standard inflationary scenario), and therefore in this case the energy-scale of inflation does not determine the amplitude of the perturbations, rather $|\mR_k|$ is determined by $\sigma$.  Since the same is true for tensor perturbations, it may at first appear that this scenario will generate a tensor-to-scalar ratio that is too large.  However, there are two possibilities that can produce a sufficiently small tensor-to-scalar ratio: (i) if there are different parameters $\sigma$ and $\sigma_T$ in the modified dispersion relations for scalar and tensor modes with $\sigma \ll \sigma_T$ \cite{Amelino-Camelia:2013tla}, or (ii) if the background space-time undergoes slow-roll inflation, as shall be explained next.

Finally, there are likely other types of background evolution that could sufficiently redshift the cosmological perturbations while preserving their scale-invariance.  Whether any are as simple to implement as inflation is a question left for future work.

\section{Modified Dispersion Relations and Slow-roll Inflation}
\label{s.obs}

In the two previous sections, we considered a specific type of modified dispersion relation---motivated by dimensional reduction arguments---on the grounds that the resulting vacuum quantum state is already scale-invariant.  The interesting converse to the case where scale-invariance appears no matter the background evolution is the case where scale-invariance appears no matter the modifications to the dispersion relation.  As shall be shown in this section, under slow-roll inflation and assuming vacuum initial conditions, cosmological perturbations become scale-invariant when they exit the Hubble radius for any power law dispersion relation.

To be specific, in this section (as in Sec.~\ref{s.infl}) we shall assume that $H \gtrsim 10^{-4} / \sqrt{G \hbar}$ (the specific minimal value for $H$ will depend on the particular modified dispersion relation), and add the slow-roll condition.  Thus, this scenario is very different from the standard slow-roll inflationary scenarios where the Hubble rate must be much lower.  This large Hubble rate ensures that the term with the higher power of $k$ in the modified dispersion relation is the dominant one when the modes exit the Hubble radius and freeze.  For this reason, the calculations in this section will give different results than other studies of modified dispersion relations in the usual slow-roll inflationary models with a much lower Hubble rate, since in that case the dominant term in the modified dispersion relation at the time of horizon-exit is $k^2$.

In slow-roll inflation, the evolution of the background space-time is characterized by the two slow-roll parameters
\be
\ep = -\f{\dot{H}}{H^2}, \qquad
\eta = 2 \ep - \f{\dot\ep}{2 H \ep},
\ee
which satisfy the conditions $0 < \ep \ll 1$ and $|\eta| \ll 1$, and parametrize departures from exact de Sitter space-time.  Since in slow-roll inflation the Hubble rate is nearly constant, for a short period of time near some arbitrary conformal time $\tau_o$ the scale factor can be approximated by (see, e.g., \cite{Agullo:2009zi})
\be \label{a-appr}
a(\tau) = \f{|\tau|^{-(1+\ep)}}{H_o |\tau_o|^{-\ep}},
\ee
where $H_o$ is the Hubble rate evaluated at $\tau_o$.   If $\tau_o$ is chosen to be close to the horizon-crossing time of the perturbation Fourier modes of interest then this approximate solution is sufficient to determine the long-wavelength spectrum of the perturbations, since (given the short-wavelength solution) the long-wavelength solution for the perturbations is determined by the background evolution at horizon-crossing.

Another important relation is, to first order in $\ep$ and $\eta$,
\be
\f{z''}{z} = \f{(a \sqrt\ep)''}{a \sqrt\ep} = \f{2 + 9 \ep - 3 \eta}{\tau^2};
\ee
this can be integrated to the same level of approximation as for $a(\tau)$ in \eqref{a-appr}, giving \cite{Agullo:2009zi}
\be
z(\tau) = \f{\sqrt{\ep_o}}{\sqrt{4 \pi G} \, H_o} \,
\f{|\tau|^{-(1 + 3 \ep - \eta)}}{|\tau_o|^{-(3 \ep - \eta)}},
\ee
where $\ep_o$ is the value of the slow-roll parameter $\ep$ at the conformal time $\tau_o$.

To simplify calculations, we will assume that only one term of the order $k^{n+2}$ is relevant in the modified dispersion relation, in which case the modified Mukhanov-Sasaki equation has the form
\be \label{diff-kn}
v_k'' + \f{\sigma^n k^{n+2}}{a^n} v_k - \f{2 + 9 \ep - 3 \eta}{\tau^2} v_k = 0,
\ee
where $n$ is any positive real number.  Note that this captures any modified dispersion relation where the evolution has been adiabatic by choosing $n$ as the dominant term at horizon-crossing.  (If the evolution of the background is not adiabatic or if the modified dispersion relation $E(k)$ is not monotonic in $k$, then the WKB approximation that the above argument is based upon may well fail, in which case other terms cannot be ignored and a more careful analysis will be necessary, along the lines of what is done in \cite{Martin:2000xs, Brandenberger:2000wr, Joras:2008ck}.)

As before, the vacuum short-wavelength solution (when $z''/z \ll \sigma^n k^{n+2} / a^n$) can be approximated by the WKB solution,
\be
v_k = \sqrt \f{\hbar}{2 \, \om_k} \, e^{-i \int d\tau \, \om_k},
\ee
where now $\om_k = \sqrt{\sigma^n k^{n+2} / a^n}$, and the long-wavelength solution (when $z''/z \gg \sigma^n k^{n+2} / a^n$) is given by
\be
v_k = A_k |\tau|^{-1 - 3 \ep + \eta} + B_k |\tau|^{2 + 3 \ep - \eta},
\ee
where the prefactors are determined by imposing continuity in $v_k$ and $v_k'$ at horizon crossing.  In an expanding slow-roll inflationary space-time, the first term will quickly dominate and therefore it is only necessary to calculate $A_k$.

The horizon crossing time $\tau_k$ for the $k^{th}$ Fourier mode is when $z''/z = \sigma^n k^{n+2} / a^n$, giving
\be \label{horizon-tau}
|\tau_k| = \left( \f{2^{1/n}}{H_k^{1+\ep} \sigma} \right)^{\tf{n}{n+2+n\ep}}
\cdot k^{-1 + \tf{n \ep}{n+2}}.
\ee
Then, a short calculation shows that the matching conditions at the horizon crossing time imply that, keeping only the dependence on $k$,
\be
A_k \sim k^{-\tf{3}{2} - 3 \ep + \eta + \tf{3 n \ep}{2(n+2)}}.
\ee

From this, it is straightforward to calculate the scalar power spectrum of the co-moving curvature perturbations after they have exited the horizon (dropping the slow-roll parameters in the exponents of $H$, $\sigma$ and $\tau_o$),
\be
\Delta_\mR^2(k) \approx \f{2^{\tf{4-n}{2(n+2)}}G\hbar}{\pi}
\cdot \f{H^2 (H \sigma)^{\tf{-3n}{n+2}}}{\ep} \cdot
k^{n_s - 1},
\ee
where the scalar spectral index is
\be
n_s - 1 = -6 \ep + 2 \eta + \f{3 n \ep}{n + 2},
\ee
showing that the perturbations have a nearly scale-invariant spectrum.

Note that the amplitude is determined by a combination of $\sigma, H$, and $\ep$.  For $n=0$, the results of standard single-field slow-roll inflation are recovered, while for $n=4$ the results are exactly those obtained in \cite{Magueijo:2008yk} (and reviewed above in Sec.~\ref{s.mod}).  For other values of $n$, all three parameters contribute to the amplitude.

It is important to keep in mind that the Hubble rate $H$ is bounded below by \eqref{bound-H}.  Observations constrain this model more strongly if $n<4$ because in that case the Hubble rate appears in the numerator of the power spectrum.  On the other hand, if $n>4$ then a large Hubble rate (compared to $\sigma^{-1}$) actually suppresses the amplitude of the scalar perturbations.

In addition, the required value of $\sigma$ for the model to give the observed amplitude will depend on $n$.  For example, for $n=2$, under the assumption $\epsilon \sim 10^{-2}$ and parametrizing $H \sim 10^\alpha / \sqrt{2} \sigma$ (with $\alpha \gtrsim 1$ for the condition \eqref{bound-H} to be satisfied), the amplitude is $\Delta_\mR^2(k) \sim 10^{-9}$ for $\sigma \sim 10^{(22+\alpha)/4} \sqrt{G\hbar}$.  In particular, the $n=1$ case is ruled out since in that case the required value for $\sigma$ is incompatible with observational constraints \cite{Ackermann:2009aa}.

Note that corrections from modified dispersion relations contribute a term that gives a slight blue tilt to the spectrum if this term is large enough.  Therefore, in order to generate a red tilt as is observed in the CMB, in this scenario it is necessary to assume that $n$ is sufficiently small or that $\eta$ is negative and sufficiently larger in amplitude than $\ep$, depending on the value of $n$.

Another interesting result is that if $n=4$ (as suggested by dimensional reduction arguments \cite{Horava:2009if, Sotiriou:2011aa}), then $n_s - 1 = -4 \ep + 2 \eta$.  Since $\eta = 2 \ep - \dot\ep / 2 H \ep$, a red tilt can obtained for positive $\dot \ep$, even when $n$ is exactly 4.  Therefore, it is not necessary to assume a long transient period between $n=4$ at very high energies and $n=0$ at low energies as proposed in \cite{Amelino-Camelia:2013tla} in order to obtain a red tilt in the scalar spectral index: instead it is enough to have an equation of state that is increasing in time.

Finally, this calculation can be repeated for tensor modes $h_k$, with the only difference that the time-dependent potential is $a''/a$ rather than $z''/z$, with the Mukhanov-Sasaki equation with a modified dispersion relation being
\be
\mu_k'' + \f{\sigma_T^n k^{n+2}}{a^n} \mu_k - \f{2 + 3 \ep}{\tau^2} \mu_k = 0,
\ee
where $\mu_k$ is related to the tensor perturbation via $h_k = \mu_k / a$.  Following the same procedure of matching the short-wavelength quantum vacuum solution
\be
\mu_k = \sqrt \f{\hbar}{2 \, {\om_T}_k} \, e^{-i \int d\tau \, {\om_T}_k},
\ee
(where ${\om_T}_k = \sqrt{\sigma_T^n k^{n+2} / a^n} \, $) to the long-wavelength solution
\be
\mu_k = \tilde{A}_k |\tau|^{-1 - \ep} + \tilde{B}_k |\tau|^{2 + \ep}
\ee
at the horizon crossing time \eqref{horizon-tau} (with $\sigma$ again replaced by $\sigma_T$) gives the form of the tensor modes at late times, from which the power spectrum of the tensor perturbations can be calculated.  The result of this calculation is that the predicted power spectrum for the tensor modes is
\begin{align} \label{tens-spec}
\Delta_h^2(k) &= 64 \pi G \f{k^3}{2 \pi^2} |h_k|^2 \nn \\ &
\approx 16 \, \f{2^{\tf{4-n}{2(n+2)}} G\hbar}{\pi}
\cdot H^2 (H \sigma_T)^{\tf{-3n}{n+2}} \cdot
k^{n_t},
\end{align}
where the tensor spectral index is
\be
n_t = - 2 \ep + \f{3 n \ep}{n+2},
\ee
showing that the tensor modes also have a nearly scale-invariant power spectrum (and in fact an exactly scale-invariant spectrum for $n=4$).

The tensor-to-scalar ratio is given by
\be \label{pred-r-gen}
r = \f{\Delta_h^2(k)}{\Delta_\mR^2(k)} = 16 \, \ep \, \left( \f{\sigma}{\sigma_T} \right)^{\tf{3}{n+2}}.
\ee
The simplest choice for $\sigma_T$ is to set it equal to $\sigma$.  With this choice, the tensor-to-scalar ratio is the same as for standard slow-roll inflation:
\be
r = 16 \, \ep.
\ee
This shows that, for the case when the background space-time undergoes slow-roll inflation, it is possible to have $\sigma_T = \sigma$ and generate tensor perturbations with a sufficiently small amplitude that is consistent with observations.  (Of course, it is also possible that $\sigma_T \neq \sigma$, in which case the predicted value of $r$ is given by the more general result \eqref{pred-r-gen}.  Note that gravitational Cherenkov radiation would appear if gravitational waves travel slower than the matter fields; this effect essentially rules out the possibility that $\sigma_T < \sigma$ \cite{Moore:2001bv, Kiyota:2015dla}.  Interestingly, the case $\sigma_T > \sigma$ remains possible, in which case the amplitude of $r$ would be suppressed \cite{Amelino-Camelia:2013tla}.)

Note that we have also assumed that the modification to the power of $k$, denoted by $n$, in the modified dispersion relations for scalar and tensor perturbations are the same.  It is easy to allow for a different value of $n$ for tensor modes simply by replacing $n$ by $n_{(T)}$ in \eqref{tens-spec}.

An interesting next step would be to go beyond linear perturbation theory and study non-Gaussianities to check what effect modified dispersion relations may have in that setting, and whether the predictions are consistent with the latest observations.  If there are different modifications to the dispersion relations for the scalar and tensor modes, this could generate new effects in three-point functions combining scalar and tensor modes, although these particular three-point functions would likely have a very small amplitude.

\section{Discussion}
\label{s.disc}

Modified dispersion relations capture some important features of the spontaneous dimensional reduction in the ultraviolet that a number of theories of quantum gravity appear to predict.  Modified dispersion relations can play an important role in a cosmological context, and in fact for a particular type of modified dispersion relation, $E^2 = k^2 + \sigma^4 k^6$, the vacuum quantum state of the shortest wavelength cosmological perturbations is already scale-invariant.  This suggests that modified dispersion relations may be the ultimate source of the observed scale-invariance in the CMB.  However, it is clear that modified dispersion relations alone are not sufficient: it is necessary to redshift these short-wavelength perturbations to cosmological scales without ruining their scale-invariance.

While modified dispersion relations were initially introduced as an alternative to inflation, on their own---i.e., without a new cosmological era before the standard radiation-dominated one---they cannot explain the scale-invariance observed in the cosmic microwave background for the reason that, without this extra cosmological era, their predictions today would concern wavelengths of the order of $\, \sim 1$ meter, not cosmological scales.  To reach cosmological scales, an additional period of expansion before the beginning of the radiation-dominated era which redshifts the perturbations by a factor of $\sim 10^{26}$ is required.

This problem can be solved by adding an inflationary period (which need not be slow-roll).  While the resulting cosmology is an extension of inflation rather than an alternative to it, the predictions of modified dispersion relations combined with inflation (with a large Hubble rate) can now explain the temperature anisotropies at the wavelengths observed in the CMB and in addition these predictions are slightly different to those of standard inflationary scenarios.  Of course, it may be possible to find one or more other types of cosmological expansion that can provide the necessary redshifting of the perturbations instead of inflation, in which case these would provide viable alternatives to inflation based on modified dispersion relations; this possibility is left for future work.

Importantly, inflation with a sufficiently large Hubble rate not only redshifts cosmological perturbations, but in addition the perturbations freeze sufficiently early so that their scale-invariance is preserved if they exit the Hubble radius at a time when the $k^6$ term in the modified dispersion relation is the dominant term.  This requires the Hubble rate to be greater than $\sigma^{-1}$ at the horizon-crossing time.

Finally, we also considered a generic power law modified dispersion relation $E^2 \sim \sigma^n k^{n+2}$ in the case when the background FLRW space-time is undergoing slow-roll inflation.  Assuming the cosmological perturbations are initially in their vacuum state, the resulting power spectrum after the perturbations exit the Hubble radius is nearly scale-invariant, no matter the modified dispersion relation, and for $n \ge 2$ it is possible to choose a phenomenologically allowed value of $\sigma$ to obtain the observed amplitude.  In addition, a small tensor-to-scalar ratio is predicted, even if the modifications to the dispersion relations of scalar and tensor modes are identical.  (For a background evolution far from slow-roll inflation, a small tensor-to-scalar ratio can only be obtained if the modified dispersion relations for the scalar perturbations and the tensor perturbations are different \cite{Amelino-Camelia:2013tla}.)

The slow-roll parameter $\eta$ affects the scalar spectral index no matter the power in the modified dispersion relation.  This is important because, for the modified dispersion relation $E^2 \sim \sigma^4 k^6$, the cosmological perturbations in their vacuum quantum state are exactly scale-invariant and therefore this result shows that a red tilt can be generated at large scales by FLRW space-times where the effective equation of state evolves with time.  To be specific, a red tilt is generated if the effective equation of state is increasing at the time of horizon-crossing, while a blue tilt is obtained if the effective equation of state is decreasing at the time of horizon-crossing.

It is interesting that a lot of the predictions of modified dispersion relations in a background undergoing slow-roll inflation---including in the limit that the modifications to the dispersion relation are important---appear to be qualitatively very similar to those of standard slow-roll inflation.  The main difference is in the scalar and tensor spectral indices which contain corrections coming from the modified dispersion relations.  In particular, these modifications lead to a violation of the consistency relation of standard single-field inflation $r = -8 n_t$ to
\be
r = \f{- \, 8 n_t}{1 - \tf{3n}{2n+4}},
\ee
in the case that the modified dispersion relations are identical for scalar and tensor perturbations.  (Note that this relation is singular for $n=4$ as in that case $n_t=0$ while $r = 16 \ep$.)  This modification to the consistency relation of standard inflation provides a potential observational test for modified dispersion relations in slow-roll inflation.


\acknowledgments

We would like to thank Iv\'an Agull\'o for helpful discussions.
This work was supported by a grant from the John Templeton Foundation. The work of SB is supported by DAAD and by the foundation Blanceflor Boncompagni-Ludovisi, n\'ee Bildt.

\raggedright

\end{document}